\documentstyle{article}
\def\1ad{\mbox{\normalsize $^1$}}
\def\2ad{\mbox{\normalsize $^2$}}
\def\3ad{\mbox{\normalsize $^3$}}
\def\4ad{\mbox{\normalsize $^4$}}
\def\5ad{\mbox{\normalsize $^5$}}
\def\6ad{\mbox{\normalsize $^6$}}
\def\7ad{\mbox{\normalsize $^7$}}
\def\8ad{\mbox{\normalsize $^8$}}
\def\makefront{\vspace*{1cm}\begin{center}
\def\newtitleline{\\ \vskip 5pt}
{\Large\bf\titleline}\\
\vskip 1truecm
{\large\bf\authors}\\
\vskip 5truemm
\addresses
\end{center}
\vskip 1truecm
{\bf Abstract:}
\abstracttext
\vskip 1truecm}
\def\defop#1{\expandafter
\def\csname#1\endcsname{\mathop{\rm#1}\nolimits}}
\defop{tr}
\catcode`\@=11
\def\slash{\@ifnextchar[{\@slash}{\@slash[\z@]}}
\def\@slash[#1]#2{
\setbox\z@\hbox{$#2$}\@tempdima\wd\z@\box\z@%
\@tempdimb#1 \advance\@tempdimb-\@tempdima \kern\@tempdimb
\hbox to\@tempdima{\hss\@makeslash\hss}}
\def\@makeslash{$/$}
\catcode`\@=12
\newcommand{\beq}{\begin{equation}}
\newcommand{\eeq}{\end{equation}}
\newcommand{\be}{\begin{eqnarray}}
\newcommand{\ee}{\end{eqnarray}}
\newcommand{\dd}{{\mathrm d}}
\newcommand{\e}{{\mathrm e}}
\newcommand{\de}{\partial}

\newcommand{\Dslash}{{\slash{D}}}
\newcommand{\half}{{1 \over 2}}
\newcommand{\ua}{{\underline{\alpha}}}

\begin {document}
\def\titleline{
Holographic Renormalization Group and 
\newtitleline
Fermionic Boundary Data
}
\def\authors{
J.~Kalkkinen\1ad and D.~Martelli\2ad
}
\def\addresses{
\1ad
Institut Henri Poincar\'e \\
11, rue Pierre et Marie Curie\\
F--75231 Paris, France 
\newtitleline  
\2ad
Department of Physics\\
Queen Mary and Westfield College,
University of London\\
Mile End Road\\
London, E1 4NS, UK
}
\def\abstracttext{
We discuss Holographic Renormalization Group equations 
in the presence of fermions and form fields in the bulk. 
The existence of a holographically dual quantum field theory 
for a given bulk gravity
theory imposes consistency conditions on the ranks of the form fields, 
the fermion - form field couplings, 
and leads to a novel Ward identity.
}
\large
\makefront

\section{Introduction}

In theories with diffeomorphism invariance the Poincar\'e generators are gauged, and 
a space-time point is not a gauge invariant concept. 
It becomes therefore at least very difficult to
define local observables. The only known way around this problem is provided by  
Holography \cite{holography}: According to this principle
a theory with diffeomorphism invariance 
should be describable in terms of 
a dual local quantum field theory defined on the boundary of the bulk space-time. 
This conjecture was originally obtained from considerations involving black hole entropies: 
Rather than growing 
as a function of the volume of the system, they seemed to behave proportionally to 
the area of a bounding surface.
A precise connection between strings and confining quantum field theories was already anticipated 
by Polyakov, cf.~eg.~\cite{polyakov};
The AdS/CFT conjecture \cite{maldacena} makes these ideas very concrete, even providing an exact relationship between
the correlators of the quantum field theory on the boundary and the gravity theory in the bulk. 

By holographic duality  we mean here that the following rather specific requirement hold: 
Given a classical diffeomorphism invariant action $S[\phi]$ on a space-time $M$ 
there should exist a dual quantum field theory with a generating functional $W[J]$ which 
coincides with the bulk action
\be
\e^{-S[\phi_{\mathrm{cl}}]} &=& \e^{-W[J]}
\ee
when the classical action is evaluated on solutions $\phi_{\mathrm{cl}}$ 
that coincide with the currents on the boundary: $\phi_{\mathrm{cl}}|_{\partial M} = J$. 
In this way the local boundary quantum field 
theory manages to assemble all classical nonlocal dynamics of 
the bulk theory. Conversely, the classical bulk theory yields now relatively
easily strong nonperturbative results on the QFT side.

The problem we want to attack now (following \cite{julkaistu}, see also \cite{konsistenssi}) 
is to understand 
what theories -- both in the bulk and on the boundary -- may have a holographic dual. 
There are some indications that supersymmetry, for instance, might necessarily play  a role; 
another characteristic that might arise naturally is the conformal symmetry also outside the fixed point description.
These characteristics, if present, would naturally be very strong, rather undesired
constraints on the validity of holography. 

We shall choose to use the very specific description of
holographic renormalization group flows due to de Boer, Verlinde and Verlinde \cite{dBVV}: Another motivation 
for these investigations is to find out under what conditions 
a similar flow equation to theirs can be formulated 
in the presence of spin and form fields. Finally, on the course of these 
investigations we shall encounter interesting phenomena connected to 
the interplay between the bulk constraints and the boundary Ward identities.

\section{Regularization}

As Holography turns the problem of finding 
the generating functional of a quantum 
field theory into the problem of solving the classical action given 
a set of initial data, the appropriate 
framework is the Hamilton--Jacobi theory. 
One finds such a canonical transformation that  
the new phase space coordinates are constants of motion. 
The generating functional $F[q]$ of the canonical transformation must solve 
simultaneously all the constraints ${\cal G}[p,q]$ of the classical theory using 
\be
p=\frac{\delta F[q]}{\delta q}~. \label{pqF}
\ee
The function $F[q]$ 
is then the classical action evaluated at some given (radial) time $r$ 
for fixed boundary values $q(r)$.

From the QFT point of view most of the bulk constraints are Ward identities. 
An exception is naturally the constraint that generates 
radial translations (Hamiltonian). It turns out that, in a generic theory containing gravity and scalars, 
this constraint equation can be written symbolically as
\be
(F,F) &=& L~.
\ee
The bracket produces first functional derivatives of the arguments w.r.t.\ the 
various fields on the boundary; the functional $L$ is the potential part of the original 
bulk action.

An action that solves these equations will typically diverge at some value 
of the transversal coordinate $r=r_*$, and must therefore be regularized by 
subtracting from it a suitable counter term Lagrangian $S_{\mathrm{ct}}$
\be
W=F-S_{\mathrm{ct}}~.
\ee
The presence of these divergences in the classical theory is expected, as they correspond 
on the QFT side to the renormalization group fixed points \cite{renor}.
We should therefore first solve
\be
(S_{\mathrm{ct}},S_{\mathrm{ct}}) &=& L - L_{\mathrm{finite}}~,\label{counterterm}
\ee
so that the remaining piece satisfies
\be
(S_{\mathrm{ct}},W) = -\half (W,W) + L_{\mathrm{finite}}~.
\ee
Loosely speaking, this can be written in the form
\be
\beta \frac{\partial W}{\partial J}  = \half (W,W) + L_{\mathrm{finite}}~,
\ee
where $\beta = \frac{\partial S_{\mathrm{ct}}}{\partial J}$ 
plays the role of the beta functions. 
We shall call this equation the Callan--Symanzik equation. Notice that this procedure is
not unique, but two solutions may differ by finite parts. When solving the Eq.~(\ref{counterterm}) 
for $S_{\mathrm{ct}}$, one should give the ansatz as a Laurent expansion around $r=r_*$, and solve 
the equation order by order.

\section{Model}

We will consider the most general two-derivative action 
with quadratic fermion couplings consistent with gauge symmetry 
\be
S &=& \frac{1}{\kappa^2} \int \dd^{d+1} x \sqrt{g} \left(  \tilde{R} 
- 2\eta \tilde\nabla\cdot\tilde\nabla_n n + 2\eta 
\tilde\nabla \cdot (n \tr K)  - \kappa^2 \Lambda \right)  \label{sI}  \\
 &+&  \int \dd^{d+1} x \sqrt{g} \left( \frac{1}{2 \lambda^2 } F_A F^A + F_A ~ \bar\psi \zeta \Gamma^A \psi 
+ \half{\bar\psi}
M \Dslash \psi - \half(\Dslash{\bar\psi})M\psi +  \bar{\psi} Z_A \Gamma^A \psi 
 \right) \nonumber ~. \nonumber
\ee
The field strength $F = \dd A$ is an Abelian $p$-form; 
There can be arbitrarily many fermion flavours, 
but we always suppress the index that would distinguish them. 
(For further notation, see Ref.~\cite{julkaistu}.)
This action encaptures, and generalizes, many 
interesting features of the effective superstring actions. For instance, $F$ 
could be thought of as a Ramond--Ramond field.

Finding the Hamiltonian formulation of this theory is straight forward though laborious. 
The result includes first class constraints that generate translations 
in the bulk, Lorentz rotations of the spinors and 
gauge transformations of the form fields. Due to the presence of spinors also 
a second class constraint between the spinorial momenta and coordinates arises. 
This means that one needs a Lorentz invariant way of dividing the fermionic phase space 
in coordinate and momentum directions; the boundary conditions are then imposed only on the coordinate spinors.

There are essentially three ways to proceed: The one already applied in 
AdS/CFT correspondence uses the split into positive and negative chirality 
spinors \cite{spinors}. One may also divide complex spinors in real and imaginary parts 
provided one can impose a Majorana condition. If there are several fermion 
flavours, one may pair them and use the arising symplectic flavour 
structure as the symplectic phase space structure. 
In addition, by breaking Lorentz invariance explicitly e.g.~using 
the form field we have in the model one may try to impose generalized 
Majorana and chirality conditions. Here we shall be mostly 
involved with the chiral splitting, though we have worked out the 
details of the other splittings as well, and the results are similar.

In terms of the eigenstates of the chirality operator associated to the normal direction of the boundary 
the fermionic kinetic term separates into
\be
- \sqrt{\eta \hat{g}} ( \bar\psi_- M \dot\psi_- +  \dot{\bar\psi_+}  M \psi_+ )  
- \partial_t {G}  -\half \bar\psi_- \partial_t ( \sqrt{\eta \hat{g}} M)~  \psi_- 
 - \half \bar\psi_+ \partial_t ( \sqrt{\eta \hat{g}} M)~  \psi_+ \label{kinfermi}~.
\ee
Here the total time-derivative term $\de_t G$ 
should actually be simply subtracted from the action, 
as argued in \cite{spinors}. 
Generally speaking the reason for this procedure is that the generating functional arises also 
on the bulk side, eventually, from a path integral propagator, defined naturally 
in Hamiltonian language. 
There will also arise terms that involve a time 
derivative of the metric: This will naturally 
change the gravitational momenta, and 
it turns out to be necessary to shift them by
\be
\pi^{ij} &\longrightarrow& \pi^{ij} - \half \hat{g}^{ij} G~. 
\ee
The remaining terms then fix the 
symplectic structure of the fermion phase space.

\section{The Holographic Callan--Symanzik Equation}
\label{bracket}

Having split the fermion phase space we 
can now write down the Hamilton--Jacobi equations for the full system. 
It turns out that, provided there are no marginal operators $Z$ present 
in the bulk and that the rank of the tensor field $p$ is {\em odd},
the Hamilton--Jacobi equation originating from the constraint that generates translations 
transversal to the boundary does indeed 
take the  generally expected form 
\be
(F,F) = {L}~\label{bracketeq}~,
\ee
where now 
\be 
(F,F) &=& (F,F)_g + (F,F)_A + (F,F)_\varphi~,
\ee
and the right-hand side of  (\ref{bracketeq}) is 
\be
{L} &=& \sqrt{\hat{g}} \left( \frac{1}{\kappa^2}  R - \Lambda 
 + \frac{1}{2 \lambda^2 } F_{\hat{A}} F^{\hat{A}} + 
F_{\hat{A}}  \bar\varphi \zeta \Gamma^{\hat{A}} \varphi + \half  {\bar\varphi}
M \hat{\Dslash} \varphi - \half (\hat{\Dslash}{\bar\varphi})M\varphi
 \right) ~.
\ee
It is useful to define the following operators:
\be
{\cal D} &=& \half \bar\varphi \frac{\delta}{\delta \bar\varphi} 
-  \half\frac{\delta}{\delta \varphi} \varphi \label{bigd} \\
{\cal D}^{ij} &=&  \frac{\delta}{\delta g_{ij}} - \half g^{ij} {\cal D} \\
{\cal D}^{\hat{A}} &=& \frac{\delta}{\delta A_{\hat{A}}} +
\bar\varphi \zeta M^{-1} \Gamma^{\hat{A}} \frac{\delta}{\delta \bar\varphi} + 
\frac{\delta}{\delta \varphi} M^{-1}\zeta \Gamma^{\hat{A}} \varphi ~.
 \ee
Had we also considered $p$ even, 
the last equation would have been different: Then the operator ${\cal D}^{\hat{A}}$ 
would have contained terms with either no or two derivatives w.r.t.~the fermion fields. 
Derivatives act from the left, but not on fields included in the same operator. 
Now the brackets can be written easily 
\be
(F,H)_g &=& \frac{-\eta \kappa^2}{\sqrt{\hat{g}}} (g_{il}g_{jk} - \frac{1}{d-1} g_{ij} g_{kl})~   
({\cal D}^{ij} F)~ ({\cal D}^{kl} H) \\
(F,H)_A &=& \frac{\eta \lambda^2}{2 \sqrt{\hat{g}}}~  
({\cal D}^{\hat{A}} F)~ ({\cal D}_{\hat{A}} H) \\
(F,H)_\varphi &=& \frac{-\eta}{2 \sqrt{\hat{g}}}~  \left( 
\frac{\delta F}{\delta \varphi} M^{-1} \hat{\Dslash} \frac{\delta H}{\delta 
\bar\varphi} 
- (\hat{\Dslash} \frac{\delta F}{\delta \varphi}) M^{-1} \frac{\delta H}{\delta \bar\varphi}
 \right)~.
\ee
The reason for the fact that also the fermionic momenta give rise to a bracket 
is easily seen in the case for Weyl fermions: The chiralities of the coordinates 
and the momenta are such that if we insert any operator even in Clifford matrices 
between them,  $\bar\pi {\cal O}_{\mathrm{even}} \varphi$, 
the result is nontrivial. Similarly, the nontrivial results
for odd operators arise from insertions between either two coordinates, 
$\bar\varphi {\cal O}_{\mathrm{odd}} \varphi$,   or two momenta, $\bar\pi {\cal O}_{\mathrm{odd}} \pi$. 
An analogous structure arises also for Majorana fields.

Including higher order interactions could be a problem, because they would 
introduce higher powers of momenta, which would spoil the basic form of the 
brackets. 
However, we have seen above  an encouraging rearrangement of terms,  where
the structure of the theory solves a similar problem. For instance, the form field kinetic 
term absorbs some four fermion couplings in the expression 
$({\cal D}^{\hat{A}} F)^2$. We can indeed view the fermionic additions 
in ${\cal D}^{\hat{A}}$ as a covariantization of 
the flat derivative with respect to the form field $\hat{A}$.

\section{Constraints and Ward identities}

In addition to the constraint that essentially 
generates the (radial) time evolution treated above, 
we have to solve also the rest of the first class constraints that generate,
generally speaking, gauge symmetry, 
diffeomorphism invariance and local Lorentz symmetry on the boundary.
The reason why this might not be straight forward is that, in the bulk, these constraints 
generate symmetry transformations through Dirac brackets. 
Due to the ansatz (\ref{pqF}) in Hamilton--Jacobi formalism they will
act differently on $F$.
As far as the Lorentz and the gauge symmetries are concerned the geometrically expected
actions are obtained. In case the of diffeomorphism invariance some additional 
constraints arise.

For instance, the fact that $A_{\hat{A}}$ enters the generating functional 
$F$ only 
through its field strength is sufficient to solve the constraint.
This simply reflects the fact that the boundary theory must
have the same gauge symmetry as the bulk theory. A similar situation 
prevails as far as the Lorentz constraint is concerned: 
It guarantees 
local Lorentz invariance on the boundary. Choosing a gauge
and Lorentz invariant $S_{{\mathrm{ct}}}$ will be enough to avoid anomalies in the boundary theory.

The situation is somewhat more involved when the constraints that guarantee
diffeomorphism invariance are considered.
Solving them we find that 
the variation
\be
{\delta}_{\chi} &=&  \nabla_k\chi^j~ L_{j\ua} \frac{\delta}{\delta L_{k\ua}} +
\chi^i \left( \partial_{[i} A_{\hat{A}]} {\cal D}^{\hat{A}} - 
  \frac{\delta}{\delta \varphi} \hat{D}_i \varphi +
 \hat{D}_i \bar\varphi  \frac{\delta}{\delta \bar\varphi} \right) \label{DI}
\ee
should annihilate the effective action. This differs from the expected  
Lie-derivative in two respects: First, the transformation of the form field 
is accompanied by a gauge transformation. Second, its action on fermions is 
modified by  
\be
\Delta_\chi \varphi &=& - \iota_\chi F_{\hat{A}}~ \Gamma^{\hat{A}} M^{-1} \zeta  \varphi  \\
\Delta_\chi \bar\varphi&=& \iota_\chi F_{\hat{A}}~ \bar\varphi \Gamma^{\hat{A}} \zeta M^{-1}
\ee
where $\Delta_\chi =\delta_\chi - {\cal L}_\chi$. 
If we want to restrict to 
theories where the boundary diffeomorphism invariance 
still prevails, we have to put this difference
to zero. 

For instance, if in  $S_{{\mathrm{ct}}}$ there is a coupling of the boundary 
fermions to an external tensor field, that field should be aligned with $F_{\hat{A}}$ in a certain way.
Further, if there is a kinetic term $\bar\varphi \hat{\Dslash} \varphi -  
(\hat{\Dslash} \bar\varphi)  \varphi$ the following restrictions should hold:
\be
{\cal L}_\chi \hat{F} &=& 0 \\
 \iota_\chi F^{i\hat{A}} \hat{D}_i \varphi &=& 0 
\ee
These are strong requirements, as they concern the 
boundary fields and not only couplings, and therefore really restrict,
from the bulk point of view, 
the set of acceptable initial conditions. The simplest way to solve them is naturally 
to exclude the kinetic terms from the action. 
They can also be solved by assuming
$\iota_\chi \hat{F}=0$: This means that the Killing isometry only changes the field 
$\hat{A}$ by generating a gauge transformation:
This would mean that there are no restrictions on the fermion fields, 
whereas the form field potential is frozen to configurations covariant 
under flows generated by $\chi$.

Considering diffeomorphism invariant terms in $S_{{\mathrm{ct}}}$, one obtains
the Ward identity
\be
{\cal L}_\chi W  + \Delta_\chi \bar\varphi \langle {\cal O}_{\bar\varphi} \rangle 
- \langle {\cal O}_\varphi \rangle \Delta_\chi \varphi 
  &=& - \Delta_\chi S_{{\mathrm{ct}}}~.
\ee
The right-hand side is the failure of the counter-terms to satisfy the full bulk diffeomorphism constraint,
while the vacuum expectation values would signal the breaking of 
Lorentz symmetry in the dual QFT. 

In order to understand the implications of these Ward identities let us 
consider as a toy model the Maxwell theory with a (decoupled) fermion:
\be
S_0 &=& -\frac{1}{4} F_{\mu\nu}^2 + \bar{\psi} \slash{\partial} \psi~.
\ee
Suppose we know, somehow, that in the full quantum theory there is a symmetry
\be
\delta A_\mu &=& \partial_\mu \lambda \\
\delta\psi &=& i \lambda \psi \\
\delta\bar{\psi} &=& -i  \bar{\psi} \lambda ~.
\ee
Our effective action fails to satisfy this symmetry 
by $\delta S_0 = i \bar{\psi} [ \slash{\partial}, \lambda] \psi$.
We can compensate by adding to the action the piece $S_1 = -  \bar{\psi} i \slash{A} \psi$.
Therefore, due to the presence of the symmetry we are forced to twist the spinor bundle by the 
line bundle, whose (Abelian) connection $A$ is. 

In the present system we see similar behaviour in that there, too, 
a symmetry principle constrains the effective action 
in a way that seems to force new complicated interactions to 
compensate the (almost) inevitable violations of the action given 
by the naive symmetry principles. The situation is much 
complicated, however, by the fact that the symmetry transformation 
involves higher rank gamma matrices, and no simple extension of a 
Dirac operator is yet available.

\section{Conclusions}

We have investigated the relationship between diffeomorphism 
invariant theories and their holographic duals, showing in particular that, 
in a theory that contains fermions, nontrivial consistency conditions arise. 
These conditions restrict, for instance, the couplings of even rank form field
field strengths to fermions in dimensions, where we have to -- or choose to -- 
consider only chiral boundary data.
Although the gauge and the Lorentz constraints did not lead to surprises, the  
Poincar\'e ones resulted in an anomalous contribution of the diffeomorphism 
Ward identity on the boundary. We have also indicated methods for finding 
counter-terms that solve these constrains exactly.


\vskip0.5cm
\noindent
{\large \bf Acknowledgements}

\smallskip
\noindent
Work supported by the CNRS and the European Commission RTN programme RTN1-1999-00116 in which 
J.K.~is associated to Imperial College, London. D.M.~is partly supported by
PPARC through SPG\#613.


\end{document}